\documentclass{article}
\usepackage{amsmath, amssymb, amsfonts}
\title{Geodesics in the conformally flat Eisenhart metric} 
\author{Hristu Culetu, \\Ovidius University, Department of Physics and Electronics, \\ Bld. Mamaia  124, 900527 Constanta, Romania\footnote{electronic address: hculetu@yahoo.com}}

\begin{document}
\numberwithin{equation}{section}
\pagenumbering{arabic}
\maketitle
\newcommand{\fv}{\boldsymbol{f}}
\newcommand{\tv}{\boldsymbol{t}}
\newcommand{\gv}{\boldsymbol{g}}
\newcommand{\OV}{\boldsymbol{O}}
\newcommand{\wv}{\boldsymbol{w}}
\newcommand{\WV}{\boldsymbol{W}}
\newcommand{\NV}{\boldsymbol{N}}
\newcommand{\hv}{\boldsymbol{h}}
\newcommand{\yv}{\boldsymbol{y}}
\newcommand{\RE}{\textrm{Re}}
\newcommand{\IM}{\textrm{Im}}
\newcommand{\rot}{\textrm{rot}}
\newcommand{\dv}{\boldsymbol{d}}
\newcommand{\grad}{\textrm{grad}}
\newcommand{\Tr}{\textrm{Tr}}
\newcommand{\ua}{\uparrow}
\newcommand{\da}{\downarrow}
\newcommand{\ct}{\textrm{const}}
\newcommand{\xv}{\boldsymbol{x}}
\newcommand{\mv}{\boldsymbol{m}}
\newcommand{\rv}{\boldsymbol{r}}
\newcommand{\kv}{\boldsymbol{k}}
\newcommand{\VE}{\boldsymbol{V}}
\newcommand{\sv}{\boldsymbol{s}}
\newcommand{\RV}{\boldsymbol{R}}
\newcommand{\pv}{\boldsymbol{p}}
\newcommand{\PV}{\boldsymbol{P}}
\newcommand{\EV}{\boldsymbol{E}}
\newcommand{\DV}{\boldsymbol{D}}
\newcommand{\BV}{\boldsymbol{B}}
\newcommand{\HV}{\boldsymbol{H}}
\newcommand{\MV}{\boldsymbol{M}}
\newcommand{\be}{\begin{equation}}
\newcommand{\ee}{\end{equation}}
\newcommand{\ba}{\begin{eqnarray}}
\newcommand{\ea}{\end{eqnarray}}
\newcommand{\bq}{\begin{eqnarray*}}
\newcommand{\eq}{\end{eqnarray*}}
\newcommand{\pa}{\partial}
\newcommand{\f}{\frac}
\newcommand{\FV}{\boldsymbol{F}}
\newcommand{\ve}{\boldsymbol{v}}
\newcommand{\AV}{\boldsymbol{A}}
\newcommand{\jv}{\boldsymbol{j}}
\newcommand{\LV}{\boldsymbol{L}}
\newcommand{\SV}{\boldsymbol{S}}
\newcommand{\av}{\boldsymbol{a}}
\newcommand{\qv}{\boldsymbol{q}}
\newcommand{\QV}{\boldsymbol{Q}}
\newcommand{\ev}{\boldsymbol{e}}
\newcommand{\uv}{\boldsymbol{u}}
\newcommand{\KV}{\boldsymbol{K}}
\newcommand{\ro}{\boldsymbol{\rho}}
\newcommand{\si}{\boldsymbol{\sigma}}
\newcommand{\thv}{\boldsymbol{\theta}}
\newcommand{\bv}{\boldsymbol{b}}
\newcommand{\JV}{\boldsymbol{J}}
\newcommand{\nv}{\boldsymbol{n}}
\newcommand{\lv}{\boldsymbol{l}}
\newcommand{\om}{\boldsymbol{\omega}}
\newcommand{\Om}{\boldsymbol{\Omega}}
\newcommand{\Piv}{\boldsymbol{\Pi}}
\newcommand{\UV}{\boldsymbol{U}}
\newcommand{\iv}{\boldsymbol{i}}
\newcommand{\nuv}{\boldsymbol{\nu}}
\newcommand{\muv}{\boldsymbol{\mu}}
\newcommand{\lm}{\boldsymbol{\lambda}}
\newcommand{\Lm}{\boldsymbol{\Lambda}}
\newcommand{\opsi}{\overline{\psi}}
\renewcommand{\tan}{\textrm{tg}}
\renewcommand{\cot}{\textrm{ctg}}
\renewcommand{\sinh}{\textrm{sh}}
\renewcommand{\cosh}{\textrm{ch}}
\renewcommand{\tanh}{\textrm{th}}
\renewcommand{\coth}{\textrm{cth}}

\begin{abstract}
The (4+1) dimensional conformally flat Eisenhart geometry is investigated in this work, stressing the contribution of the stress tensor generating its curvature. The energy-momentum tensor $T^{a}_{~b}$ is traceless and has only one nonzero component.It could be written as an anisotropic fluid with null transversal pressures and nonzero energy fluxes. The null and timelike geodesics are computed in the pure cosmological case when the Eisenhart potential energy is $V(r) = -m\omega^{2}r^{2}/2$, where $\omega$ is related to the cosmological constant $\Lambda$. Although the conformally metric is curved, the radial null geodesics $R(T)$ and $Y(T)$ are straight lines, with finite $R_{max}$ and $Y_{max}$, $Y$ being the 5th coordinate. In contrast, for a radial timelike geodesic, $Y_{max} \rightarrow \infty$ if $T \rightarrow T_{max} = 1/\omega$.
 
 \end{abstract}
 
 \section{Introduction}
The Eisenhart approach allows to find the nonrelativistic mechanics in (3+1) dimensions with a gravitational potential of the form \cite{MD}
		 \begin{equation}
		V(r) = \frac{\alpha}{r} + \beta r^{2}
 \label{1.1}
 \end{equation}
from the properties of a curved plane wave geometry in (4+1) dimensions. $\alpha$ and $\beta$ in (1.1) are constants, with  $\alpha = -Gm$, where $G$ is the gravitational constant and the spherical mass $m$ is the source of the potential. As Dunajski has noticed, when $\alpha = 0$ and $\beta>0$, $V(r)$ gives us the Hooke law. If $\beta = -\omega^{2}/2 <0$, the 2nd term in (1.1) resembles the cosmological term, with the cosmological constant $\Lambda = 3\omega^{2}/c^{2}$, where $c$ is the velocity of light \cite{MD}. In that case, the Newtonian potential (1.1) arises by taking a nonrelativistic limit of the Schwarzschild-de Sitter (S-dS) spacetime, i.e., $c\rightarrow \infty$ and $\Lambda\rightarrow 0$, but $\omega^{2} = c^{2} \Lambda /3$ is kept finite.

In \cite{MD} Dunajski investigates the impact of the cosmological term on the connection between the Equivalence Principle (EP) and the Quantum Mechanics (QM) requirements (see also \cite{DP}). He also found that, if $\alpha = 0$, the potential corresponds to the nonrelativistic limit of the dS metric. The geometry becomes conformally flat and a transformation to the Beltrami coordinates \cite{CGH, YT} leads to a nonunitary transformation between a Schrodinger wave function of a free particle and that of a particle moving in a reversed isotropic oscillator potential \cite{MD}. Dunajski and Penrose \cite{DP} discussed the role played by EP in QM in the framework of Newton-Cartan geometry. The problem is related to the measurement paradox in QM, when the gravitational effects are taken into account \cite{RP, RP2}. Moreover, combining quantum nonlocality with Newtonian twistor nonlocality, one could clarify the role of gravity in the wave function collapse.

Motivated by the role of the (4+1) dimensional Eisenhart line element with an arbitrary potential $V(\vec{x},t)$, Sec.2 is devoted to the null stress tensor which is at the origin of the curvature of the metric. In Sec.3 we search for the null and timelike geodesics in the pure cosmological case $V = -m\omega^{2}r^{2}/2$ when the Eisenhart metric has a conformally flat form. We conclude in Sec.4 with some general features of the subject studied.

From now on we shall use the geometrical units $\hbar = c = G =1$, unless otherwise specified.

\section{The Eisenhart geometry}
Coordinate transformations between different frames in the Schrodinger equation are convenient to be made by means of the Eisenhart geometry
		 \begin{equation}
	  ds^{2} = 2dy~dt +  \frac{2V(\vec{x},t)}{m}dt^{2} - d\vec{x}~d\vec{x},
 \label{2.1}
 \end{equation}
where $m$ is a constant mass, the coordinates are $(t, \vec{x}, y)$, $\vec{x}$ is the vector associated to the three spatial Cartesian coordinates and $y$ is the 5th coordinate. The authors of Refs.[1] and [2] showed that the Schrodinger equation arises from the wave equation corresponding to the metric (2.1)
		 \begin{equation}
		\Box \Phi = \frac{1}{\sqrt{g}}\frac{\partial}{\partial x^{a}}\left(\sqrt{g}~g^{ab}\frac{\partial \Phi}{\partial x^{b}}\right) = 0,
 \label{2.2}
 \end{equation}
where $\Phi (t,\vec{x},y) = exp(-imy/\hbar)\Psi (t,\vec{x})$, where the Planck constant has been introduced for completeness. The indices $a$ and $b$ run from $t$ to $y$ and $g = r^{4}sin^{2}\theta$ is the determinant of the metric. It is worth observing that $y$ and the particle mass $m$ are introduced through the phase of the wave function $\Phi$.  When the d'Alembertian (2.2) is computed one finds the time-dependent Schrodinger equation
 		 \begin{equation}
		-\frac{\hbar^{2}}{2m}\nabla^{2}\Psi + V(\vec{x},t)\Psi = i\hbar \frac{\partial \Psi}{\partial t}.
 \label{2.3}
 \end{equation}
We are interested in the pure cosmological case, when $V(\vec{x},t) = -m\omega^{2}r^{2}/2$ (in spherical coordinates) and (2.1) becomes
 \begin{equation}
  ds^{2} = 2dy~dt -\omega^{2}r^{2}dt^{2} - dr^{2} - r^{2}(d\theta^{2} + sin^{2}\theta d\phi^{2}).
 \label{2.4}
 \end{equation}
It can be shown \cite{GP} that (2.4) may acquire a conformally flat form
 \begin{equation}
 ds^{2} = \frac{1}{1 - \omega^{2}T^{2}}\left(2dY~dT - dR^{2} - R^{2}d\Omega^{2}\right),
 \label{2.5}
 \end{equation}
where $0<T<1/\omega$ and the new coordinates $T, R, Y$ are related to the old ones by \cite{MD}
 \begin{equation}
\omega T = tanh(\omega t),~~~R = \frac{r}{cosh(\omega t)},~~~Y = y - \frac{\omega r^{2}}{2}tanh(\omega t)
 \label{2.6}
 \end{equation}
and $d\Omega^{2}$ stands for the metric on the unit 2-sphere $S^{2}$. For a free particle $V(r,t) = 0$ so that $\omega = 0$, which gives a flat metric.

Our purpose now is finding the stress tensor $T^{ab}$ to be placed on the r.h.s. of Einstein's equations $G^{ab} = 8\pi T^{ab}$ for to get (2.5) (that represents a curved plane wave metric) as an exact solution. Keeping in mind that $Y$ is a null coordinate, we are looking for a stress tensor of the form
 \begin{equation}
T^{ab} = \epsilon l^{a}l^{b},
 \label{2.7}
 \end{equation}
where $l^{a} = (0, 0, 0, 0, 1)$ is a null vector ($l^{a}l_{a} = 0$). The only nonzero component is 
 \begin{equation}
T^{YY} = \epsilon = -\frac{3\omega^{2}}{8\pi} = -\frac{\Lambda}{8\pi}.
 \label{2.8}
 \end{equation}
We see that the constant $\epsilon$ arises as a negative pressure on the 5th direction. From (2.7) one obtains $T^{a}_{~a} = 0$, i.e., we are dealing with a null fluid travelling on the Y-direction. The effect of this fluid is the positive csmological constant $\Lambda = 3\omega^{2}$ in (3+1) dimensions. A similar dependence between $\Lambda$ and the 5th dimension has been put forward in \cite{HC}.

 Let us express now the above stress tensor as an anisotropic fluid using the (2.4) form of the Eisenhart geometry, by means of the following timelike velocity field 
 \begin{equation}
u^{a} = \left(1, 0, 0, 0, -\frac{1}{2}(1 - \omega^{2}r^{2})\right),~~~~u^{a}u_{a} = -1. 
 \label{2.9}
 \end{equation}
 The stress tensor could be written now as \cite{KM, HC4, FC}
 \begin{equation}
T_{~a}^{b} = (\rho + p_{t})u_{a} u^{b} + p_{t} \delta_{a}^{b}+ (p_{y} - p_{t}) n_{a} n^{b} + u_{a}q^{b} + u^{b}q_{a},
\label{2.10}
\end{equation}
	where $\rho = T_{ab}u^{a}u^{b}$ is the energy density, $p_{t}$ represents the transversal pressures ($p_{t} = p_{\theta} = p_{\phi} = p_{r})$, $n^{b}$ is a spacelike vector with $ n_{a} n^{a} = 1$ and $n_{a} u^{a} = 0$ and $q_{a}$ is the energy flux density four vector, given by
	 \begin{equation}
  q^{a} = -T^{~a}_{b} u^{b} -\rho u^{a}.
\label{2.11}
\end{equation}	
From the previous relations one finds that 
 \begin{equation}
n^{a} = \left(1, 0, 0, 0, \frac{1}{2}(1 + \omega^{2}r^{2})\right),~~~~n^{a}n_{a} = 1.
\label{2.12 }                                                                              
\end{equation}
From the Einstein equations $G^{a}_{~b} = 8\pi T^{a}_{~b}$ we obtain, by means of the software package Maple, that the only nonzero component of the energy-momentum tensor (2.10), in the spacetime (2.4), is $T^{y}_{~t} = -3\omega^{2}/8\pi$.

Inserting $a = b = r, \theta, \phi$ in (2.10) we get $T^{r}_{~r} = T^{\theta}_{~\theta} = T^{\phi}_{~\phi} = 0$, which gives us  $p_{t} = 0$ for the transversal pressures. In addition, the trace $T^{a}_{~a} = 0$ leads to $\rho = p_{y}$. The vanishing components $T^{t}_{~t},~T^{t}_{~y},~T^{y}_{~y}$ give us
\begin{equation}
\begin{split}
 T^{t}_{~t} = q_{t} - \frac{1}{2}(1 + \omega^{2}r^{2}) + q^{t} - \omega^{2}r^{2}\rho = 0,~~~ T^{t}_{~y} = 2\rho + q_{y} + q^{t} = 0,\\
 T^{y}_{~y} = \omega^{2}r^{2}\rho - \frac{1}{2}(1 - \omega^{2}r^{2})q_{y} + q^{y} = 0.
\label{2.13 } 
\end{split}                                                                             
\end{equation}
 Alternatively, from  $T^{y}_{~t}$ one obtains
\begin{equation}
T^{y}_{~t} = \frac{\rho}{2}(1 - \omega^{4}r^{4}) - \frac{1}{2}(1 - \omega^{2}r^{2})q_{t} - \frac{1}{2}(1 + \omega^{2}r^{2})q^{y} = -\frac{3\omega^{2}}{8\pi}.
\label{2.14}
\end{equation}
As far as the flux density $q^{a}$ is concerned, we get it from (2.11), in terms of energy density $\rho$
\begin{equation}
q^{a} = \left(-\rho, 0, 0, 0, \frac{\rho}{2}(1 - \omega^{2}r^{2}) + \frac{3\omega^{2}}{8\pi}\right). 
 \label{2.15}
 \end{equation}
Having the components of the energy flux, the energy density is acquired from (2.14)
\begin{equation}
 \rho =  -\frac{3\omega^{2}}{8\pi}. 
 \label{2.16}
 \end{equation}
For the invariant flux, Eq.2.15 yields $q \equiv \sqrt{q^{a}q_{a}} = |\rho| = 3\omega^{2}/8\pi$. It is worth noting that $l^{a}$ from (2.7) can be expressed in terms of $u^{a}$ and $n^{a}$. One obtains $l^{a} = n^{a} - u^{a}$.

 Let us find now the covariant acceleration $a^{b} = u^{a}\nabla_{a}u^{b}$ of an observer with the 4-velocity $u^{a}$ given by (2.9).
 We have $\dot{t} = dt/d\tau = 1,~dr/d\tau = 0,~dy/d\tau = -(1/2)(1 - \omega^{2}r^{2})$, where $\tau$ is the proper time.
 The acceleration 4-vector appears as
 \begin{equation}
 a^{b} = (0, -\omega^{2}r, 0, 0, 0).
 \label{2.17}
 \end{equation}
 One notices that an observer with $r = const.$ has a nonzero acceleration on the radial direction, given by $a^{r} = -\omega^{2}r$, albeit he/she is moving only on the 5th dimension, with $dy/dt = -(1/2)(1 - \omega^{2}r^{2})$. The radial acceleration, which is repulsive, resembles that of a static observer in de Sitter static metric.  An interpretation of this fact has its origin in the connection between $\omega^{2}$ and the cosmological constant, namely $\Lambda = 3\omega^{2}$. In other words, motion along the 5th dimension seems to generate a radial expansion.

\section{Geodesics}
For to find the geodesics in the spacetime (2.5) we write down the Lagrangean
 \begin{equation}
 2L \equiv \frac{ds^{2}}{d\lambda^{2}} = \frac{1}{1 - \omega^{2}T^{2}}\left(2\dot{Y}~\dot{T} - \dot{R}^{2} - R^{2}\dot{\theta}^{2} - R^{2}sin^{2}\theta ~ \dot{\phi}^{2} \right),
 \label{3.1}
 \end{equation}
where $\dot{T} = dT/d\lambda $, etc., and $\lambda$ is the affine parameter along geodesics (the proper time $\tau$ for timelike gedesics). We take into consideration, for simplicity, only the radial geodesics ($\theta, \phi = const.$) and $\theta = \pi/2$. In that case, the geometry depends on the time $T$ only and $Y, R$ are cyclic coordinates. We have $L = 1/2$ for timelike geodesics and $L = 0$ for null geodesics.

From the Euler - Lagrange equations
 \begin{equation}
\frac{\partial L}{\partial x^{a}} - \frac{d}{d\lambda}\frac{\partial L}{\partial \dot{x}^{a}} = 0
 \label{3.2}
 \end{equation}
one obtains
 \begin{equation}
\frac{\partial L}{\partial \dot{Y}} = \frac{\dot{T}}{1 - \omega^{2}T^{2}} = a,~~~ \frac{\partial L}{\partial \dot{R}} = \frac{\dot{R}}{1 - \omega^{2}T^{2}} = b,                  
 \label{3.3}
 \end{equation}
with $a,b$ dimensionless constants.

1.~\textbf{Null geodesics}\\
This case gives $L = 0$ and (3.1) yields
  \begin{equation}
 2\dot{Y}~\dot{T} - \dot{R}^{2} = 0.
 \label{3.4}
 \end{equation}
  From (3.3) we get
	  \begin{equation}
 \frac{dT}{1 - \omega^{2}T^{2}} = a~d\lambda,~~~ \longrightarrow T(\lambda) = \frac{1}{\omega} tanh(a\omega \lambda),                
 \label{3.5}
 \end{equation}
with appropriate initial condition. In the same manner one obtains, by means of (3.5)
	  \begin{equation}
   \frac{dR}{1 - \omega^{2}T^{2}} = b~d\lambda,~~~ \longrightarrow    R(\lambda) = \frac{b}{a\omega} tanh(a\omega \lambda).                
 \label{3.6}
 \end{equation}
Once (3.5) and (3.6) are introduced in (3.4), we have
	  \begin{equation}
   \frac{dY}{d\lambda} = \frac{b^{2}}{2a~cosh^{2}(a\omega \lambda)},~~~ \longrightarrow    Y(\lambda) = \frac{b^{2}}{2a^{2}\omega} tanh(a\omega \lambda).                
 \label{3.7}
 \end{equation}
In terms of the coordinate time $T$, (3.6) and (3.7) gives us
	  \begin{equation}
		R(T) = \frac{b}{a}T,~~~Y(T) = \frac{b^{2}}{2a^{2}}T,
 \label{3.8}
 \end{equation}
with $b\leq a$ and $R_{max} = b/(a\omega)$ because $T_{max} = 1/\omega$. The null geodesics $R(T)$ and $Y(T)$ are straight lines, where the ratio $b/a$ plays the role of the velocity of the null particle. 

2.~\textbf{Timelike geodesics}\\
 We have now $L = 1/2$, so that (3.4) becomes 
  \begin{equation}
 2\dot{Y}~\dot{T} - \dot{R}^{2} = 1 - \omega^{2}T^{2}.
 \label{3.9}
 \end{equation}
The equivalent of (3.5) and (3.6) looks now as
  \begin{equation}
	\frac{\dot{T}}{1 - \omega^{2}T^{2}} = p~~~\longrightarrow     T(\tau) = \frac{1}{\omega} tanh(p~\omega \tau)
 \label{3.10}
 \end{equation}
and
	  \begin{equation}
   \frac{\dot{R}}{1 - \omega^{2}T^{2}} = q,~~~ \longrightarrow    R(\tau) = \frac{q}{p~\omega} tanh(p~\omega \tau),               
 \label{3.11}
 \end{equation}
where $p$ and $q$ are positive dimensionless constants. As far as $R(T)$ is concerned, one finds that $R(T) = (q/p)T$, with $q<p$ and $R_{max} = q/(p\omega)$. Even a massive free particle moves on a straight line in the radial direction, $q/p$ being the velocity of the particle.  

The shape of $Y(\tau)$ changes now, compared to (3.7). After inserting the expressions of $T(\tau)$ and $R(\tau)$ from (3.10) and (3.11) in (3.9), one obtains
 	  \begin{equation}
    Y(\tau) = \frac{\tau}{2p} + \frac{q^{2}}{2p^{2}\omega} tanh(p~\omega \tau),~~~Y(0) = 0.                                                       
 \label{3.12}
 \end{equation}
In terms of the coordinate time $T$ and with the help of (3.10), the above equation yields
 	  \begin{equation}
    Y(T) = \frac{1}{4\omega p^{2}}ln\frac{1 + \omega T}{1 - \omega T} + \frac{q^{2}}{2p^{2}}T.                                                      
 \label{3.13}
 \end{equation}
It is worth noting that $Y_{max} \rightarrow \infty$ when $T \rightarrow 1/\omega$. It is a consequence of the singularity of the geometry (2.5) when $T$ approaches $1/\omega$.

\section{Conclusions}
The nonrelativistic mechanics in (3+1) dimensions can be studied by means of the Eisenhart approach, from the properties of a curved plane wave geometry in (4+1) dimensions. The Newtonian potential arises by taking a nonrelativistic limit of the Schwarzschild-de Sitter space. Moreover, the time-dependent Schrodinger equation is generated by the wave function corresponding to the 5-dimensional Eisenhart geometry with an arbitrary $V(\vec{x},t)$. 
 
We studied in this paper the pure cosmological case when $V(r) = -m\omega^{2}r^{2}/2$ and the Eisenhart metric may be written in a  conformally flat form. The stress tensor generating the curvature has only one nonzero component related to the cosmological constant $\Lambda = 3\omega^{2}$. Moreover, it is traceless and may correspond to an imperfect fluid, with null transversal pressures and nonzero energy flux densities. The radial geodesics are investigated, stressing the linear character of the functions $R(T)$ and $Y(T)$.

\end{document}